\documentclass[useAMS,usenatbib]{mn2e}

\usepackage{graphicx}
\usepackage{graphpap}
\usepackage{epsf}
\usepackage{amssymb}
\usepackage{color}

\evensidemargin=\oddsidemargin \textheight=23cm \textwidth=17.cm

\def\FIG #1 #2 [#3] #4\par{%
  \begin{figure} \begin{center}%
    \includegraphics*[#3]{#2}%
    \\
    \caption{#4}%
    \label{#1}%
  \end{center}\end{figure}%
}
\def\FIGG #1 #2 #3 [#4] #5\par{%
  \begin{figure}
    \resizebox{\hsize}{!}{
               \includegraphics*[#4]{#2}
               \includegraphics*[#4]{#3}}%
    \caption{#5}%
    \label{#1}%
  \end{figure}%
}

\def\FIGs #1 #2 #3 #4 #5 [#6] #7\par{%
  \begin{figure}
        \includegraphics[#6]{#2}
        \includegraphics[#6]{#3}\\
        \includegraphics[#6]{#4}
        \includegraphics[#6]{#5}
        \caption{#7}
        \label{#1}
  \end{figure}
}

\def\FIGss #1 #2 #3 #4 #5 #6 #7 [#8] #9\par{%
  \begin{figure}[!h]
    \begin{center}
        \includegraphics*[#8]{#2}
        \includegraphics*[#8]{#3}
        \includegraphics*[#8]{#4}
        \includegraphics*[#8]{#5}
        \includegraphics*[#8]{#6}
        \includegraphics*[#8]{#7}
        \caption{\label{#1}#9}
    \end{center}
  \end{figure}
}

\def\rfig#1{Fig.\ref{#1}}

\def\Ek{E_{\mathrm{thr}}}

\def\Ka{$K_{\alpha}$ }

\def\elk{{Paper I}}

\def\bq{\begin{equation}}
\def\eq{\end{equation}}

\def\ba{\begin{array}}
\def\ea{\end{array}}

\newcommand{\f}[2]{\frac{#1}{#2}}

\renewcommand{\la}{\langle}
\newcommand{\ra}{\rangle}

\newcommand{\avr}[1]{\la #1 \ra}

\title{Inner-shell ionization, radiative losses and thermal conductivity in young SNRs}
\author[Kosenko~ D.I.]{Kosenko~D.I.$^{1,2}$\thanks{E-mail:lisett@xray.sai.msu.ru}\\
$^{1}$Sternberg Astronomical Institute, Universitetskij pr., 13, 119992,
Moscow, Russia \\
$^{2}$Institute for Theoretical and Experimental Physics, Bolshaya
Cheremushkinskaya, 25, 117218, Moscow, Russia }

\date{Accepted 2006 March 24. Received 2006 March 23; in original form 2006 January 19}

\begin{document}

\maketitle{}
\label{firstpage}

\begin{abstract}
Self-consistent treatment of time-dependent ionization
in hydrodynamical calculations of the X-ray emission from young supernova remnants has been performed. 
The novel feature of the calculations 
is that \Ka lines from species produced by inner-shell collisional ionization are included. Parameters of the 
shocked ejecta are found from fitting the model spectrum to the observed one. 
The application of the method to Tycho SNR using the classical deflagration model W7 for the explosion
enables us to well reproduce the observed X-ray spectra and radial brightness profiles of the 
remnant.
\end{abstract}

\begin{keywords}
supernovae remnants - x-ray emission: time-dependent ionization - inner-shell ionization.
\end{keywords}

\section{Introduction}
The supernova remnant 1572 (Tycho SNR) is a convenient object for testing thermonuclear 
supernova models. The available observational data 
 \citep{exosat, ginga,xmm,chandra} makes
 it possible to test both the theory of thermonuclear supernova explosions and
gas-kinetic processes in plasma heated by the SN blast wave.
Studies of Tycho SNR \citep{itoh,brink,elka,bb1} showed that it is difficult to find a model which describes 
self-consistently its origin and evolution. Nevertheless, the increasingly accurate account for 
various physical processes in analyzing of SNR observations makes it possible to narrow the range of possible
explosion mechanisms.

As was shown in detail in \citet{elka} (hereafter  \elk),  the accurate modeling of 
young SNRs even in one-dimensional hydro calculations requires taking into account 
radiative energy losses and self-consistent calculation of plasma ionization state in every mesh zone 
at every time step. In addition, such  processes as thermal conductivity of electrons and non-Coulomb energy exchange
between electrons and ions should be included. It is very hard to treat these processes from the first principles, so 
two free parameters have been introduced. The values of these parameters can be found from comparison
with observations. In \elk\, we compared modeled emission with X-ray observations of Tycho SNR obtained by the XMM-Newton 
X-ray observatory and reported in \citet{xmm}. The data include X-ray (0.2-10 KeV) spectrum and 
azimuthally averaged profiles of X-ray brightness in silicon and iron lines (Fe~{\sc xvii}, Si~K, Fe~K). 

In \elk\, numerical simulations of SN ejecta propagated into the surrounding interstellar medium have been performed
for two models: the parametric deflagration model {\bf W7} \citep{w7}, and new 3D (the ``first principle'') model {\bf mr0} elaborated in 
Max-Plank-Institute f\"ur Astrophysik  \citep{mr}. X-ray spectra and radial brightness profiles of the 
remnant expected at an age of 430 years (the age of Tycho SNR) were computed for these models.  However,
the comparison with the XMM observations revealed noticeable discrepancies. For example, in the W7 model, 
X-ray spectrum did not reveal the presence Fe~K line and for no model the observed relative position of X-ray brightness profiles in
silicon and iron lines was obtained. However, these disagreements could not be used to reject the models
considered, since not all relevant physical processes were fully incorporated into the radiation hydro code at that time.

In later studies by \citet{granada}, an attempt  to modify the initial density profile of the SN ejecta 
was done. The outermost layers of the ejecta were made more rarefied. These modifications
showed that the density profile does not have a noticeable effect on the resulting X-ray brightness profiles 
of the remnant. However, a fine tuning of the parameters in the density distribution in the ejecta made it
possible to well reproduce the observed radial brightness profiles. In that case
most of the emission in silicon and iron lines should have been generated by the external SNR shock 
(not by the reverse shock), which seems rather unlikely  \citep{hwang}. 

When calculating X-ray emission from the remnant in  \elk, we  did not take into account 
inner-shell ionization processes from low-ionized elements.
In stationary plasma, collisional production of \Ka emission lines is suppressed due to 
a small value of the cross-section of  interaction between free and inner-shell 
electrons. Usually \Ka emission lines from low-ionization ions are produced via photoionization
processes (fluorescence). In shocked plasma  time-dependent ionization occurs and 
in a SNR the ionization time-scale is about several hundred  years 
\citep{elka}. Thus the matter behind the reverse shock front should be highly under-ionized
compared to the temperature of ionizing electrons. Hence the collisional ionization of ions 
from inner-shells is highly operational in the emission line production in a SNR.
In  \citet{vink} the role of inner-shell excitation processes to Si~$K_\alpha$ feature was investigated in SN1006.

The main goal of this paper is to assess the role of inner-shell
ionization processes in the  X-ray emission of young supernova remnants.
We calculate the hydrodynamic evolution and  X-ray emission of a young remnant
using the code {\sc supremna}, introduced in \elk, with inner-shell
ionization processes taken into account.  The example of the classical
deflagration model W7 shows that it may produce the X-ray spectrum and
radial brightness profiles very close to the ones observed from Tycho
SNR for some set of parameters. In particular, the iron \Ka line
may be pronounced in this model without invoking additional mixing in
the ejecta.

\section{Calculation of the inner-shell ionization in SNRs}
\label{calc}
To calculate the intensity of a line produced by ionization from the inner shell ($n = 1$)  of
an ion followed by the transition of an electron from level  $n$ to this vacant place,
the following formula was employed \citep{spitzer,kallm}
\bq\label{KlineInt}
{\cal J}^i_{n1} = n_i\,n_e\,\avr{\sigma v}_i\,\omega_{i. n1}\,E_{n1}\quad 
\left[\mathrm{\f {erg}{cm^3\, s}}\right]
\eq
where $n_i,\, n_e$ are number densities of ions $i$ and electrons correspondingly; 
\bq\label{fsigv}
\avr{\sigma v}_i = \int_{\Ek[i]}^\infty\sigma v\,f(v)\,dv
\eq
is the velocity averaged cross-section. Here $\Ek[i]$ is the threshold ionization energy from 
the inner-shell of  ion $i$,  $f(v)$ is the Maxwellian distribution function for electrons;
$\omega_{i. n1}$ is the fluorescence yield (photons/ionization) for the line which corrects the 
intensity for photonless transitions due to Auger electrons \citep{kaasmew,kallm};
$E_{n1}$ is the energy of the line.

Approximate formulas for the inner-shell ionization cross-sections $\sigma$
were taken from \citet{homb}. These formulas fit experimental data for a wide range 
of ions and electron energies. Each cross-section formula  was  convolved numerically
with the Maxwellian velocity distribution (formula~\ref{fsigv}). For each species dependence
$\avr{\sigma v}_i$ on temperature was numerically approximated with an accuracy of better than 1\%
and then inserted into the emission routine. Data on $\Ek[i], \omega_{i,n1}, E_{n1}$ was taken from
 \citet{kaasmew}. 

In the SNR modeling, the plasma ionization state is calculated by solving a system of kinetic equations
in each grid mesh at every time step. Assuming that inner-shell ionization processes do not affect
very much the electron production rate and thus the equation of state of the matter, the inner-shell ionization has not
been included in the kinetic equations. \rfig{xi_comp} shows ionization rates for some ions.
Ionization rates which are included in the present paper (and in \elk\,) are shown by asterisks ``*'',
the inner-shell ionization rates are presented by circles ``o''. In most cases the inner-shell ionization
rates are several times smaller  than the standard (outer-shell) ionization rates. 

We did not include inner-shell excitation processes in our method. Contribution of excitations from inner-shell is negligible compared
to normal excitations and inner-shell ionization rates. For example, formulae from \citet{mewgron} show that
the ratio of normal excitation rates to inner-shell ionization and to inner-shell excitation
for Fe~{\sc XXV} at temperature $\sim 10^8$ K is about $1:0.5:10^{-4}$.

\FIGs xi_comp rate_Ar rate_Ca rate_Fe rate_Ni [width = 0.23\textwidth]
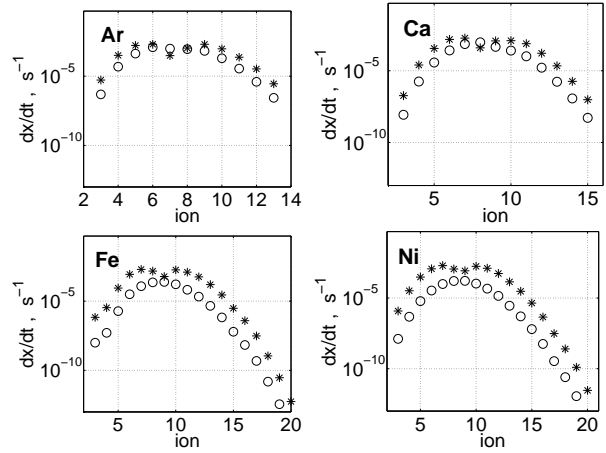
Ionization rates for Ar,Ca,Fe,Ni. Asterisks (``*'') show standard ionization rates used in SNR simulations. 
Circles (``o'') show the inner-shell ionization rates. Matter with an initial temperature of
$T_\mathrm{in} = 10^4$ K was instantly heated up to $T_\mathrm{fin} = 2\times 10^8$ K. The ionization parameter  is
$nt = 3\times 10^9\; \mathrm{s/cm}^3$.

\section{Comparison with observations}
\label{results}
The results of the calculations have been compared with the  X-ray observations of Tycho SNR performed by XMM-Newton observatory.
The data were taken from the public
available library of the observatory\footnote[1]{{\tt http://xmm.vilspa.esa.es/external/xmm\_data\_acc/xsa/}, obsID = 009621010}. 
These observations were conducted with the EPIC instrument, the spectrum was extracted from data obtained by 
the MOS1 camera.

Based on W7 explosion model, eight variants of hydrodynamical calculations have been performed for
different values of three parameters describing the thermal conductivity ($C_\mathrm{kill}$), non-Coulomb energy exchange between 
electrons and ions ($q$) and the  density of the circumstellar matter ($\rho_\mathrm{CSM}$).
Calculated spectra were converted to an {\sc xspec} \citep{xspec} table model with these three interpolation parameters.
The physical meaning of the parameters  $C_\mathrm{kill} $ and  $q$ is as follows. 

The factor $0 < C_\mathrm{kill} < 1$ describes the fraction of the thermal flux contributed to the system 
compared to the standard case (formula (7) in \elk), 
where the flux is limited only by the existence of a maximum speed (the speed of sound) for the heat carriers,
ignoring a possible decrease in the particle mean free path due to magnetic fields and plasma instabilities.
To describe the effects of collisionless energy exchange,  in \elk\, we introduced the parameter $q$ that specifies the fraction 
of artificial viscosity Q, added to the pressure of ions: $P_i = P_i(\mathrm{thermal}) + q\,Q$. 
Then for the electronic component we put $P_e = P_e(\mathrm{thermal}) + (1- q)\,Q$. If only the collisional exchange is taken into account, 
$q = 1$ and we used the standard system of equations with only ions heating of at the front.

{\bf Comparison of the spectra }

By fitting the {\sc xspec} table models to the XMM-Newton observations, we found
the values of the table interpolation parameters. Then with these values a new
hydrodynamical model was calculated. \rfig{w7sp} shows the observed spectrum (EPIC MOS1) and 
theoretical spectrum obtained for the W7 model. 
Table \ref{pars} lists the values of fitting parameters. The best-fitting values  
 for the column density $N_H$ is also shown in the table.

Inset in \rfig{w7sp} shows {\sc Fe~K} line profile in details. It is clear from the figure that modeled centroid shifted to
higher energies, compared to observable one. This discrepance indicates that the matter in our model is slightly overionized, 
thus the temperature or ionization timescale of the Tycho remnant should be lower. These arguments point to less energetic model for the remnant.

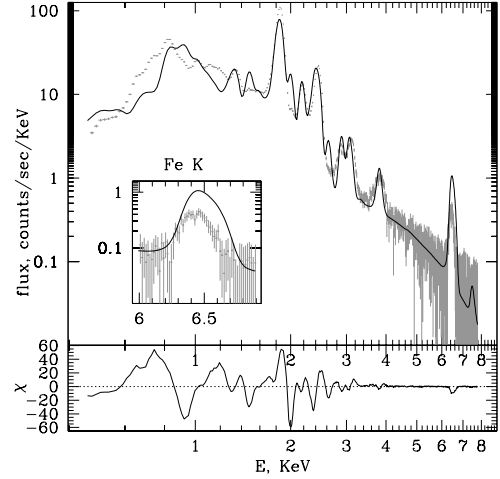
\FIG w7sp w7err1 [width=0.4\textwidth] Observed (the gray crosses) and theoretical (the black solid line) spectra of the Tycho 
SNR. Inset shows Fe \Ka line in details. 

\begin{table}
\caption{The values of the best fit parameters for Tycho SNR model}
\label{pars}
\begin{center}
\begin{tabular}{l|l|l|l|l}
\hline
\\
 $\rho_\mathrm{CSM},\,\mathrm {g/cm^{-3}}$  & q & $C_\mathrm{kill}$ & $N_H,\, \mathrm{cm}^{-2}$ \\
\hline
\\
$1.8\times10^{-24}$ & $0.93 $ & $0.0085$ & 
$3.7 \times 10^{21}$ \\ 
\hline
\end{tabular}
\end{center}
\end{table}

{\bf Comparison of the remnant's brightness profiles}

Following the scheme elaborated in  \elk,  the radial brightness profiles of the model remnant were 
constructed in Si~K, Fe~{\sc xvii}, Fe~K lines \citep{xmm}. The results are plotted in \rfig{profs}. 
The figure shows the X-ray brightness profiles for W7 model, calculated with the best-fitting
interpolation parameters (Table~\ref{pars}). These profiles are very similar to the ones found
 in \citet{xmm}.

\FIG profs w7bwprs_cmb [width=0.45\textwidth] 
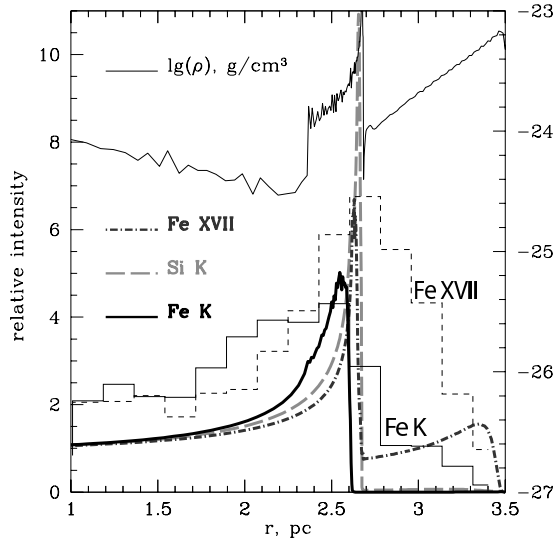
Simulated brightness profiles for the W7 model. The solid black line shows the density profile.
Step-like profiles are observed by XMM-Newton, taken from Decourchelle et al. (2001): Fe~K (solid line), Fe~XVII (dashed line).

Using the model radial profile, the distance to the remnant was evaluated.
Table \ref{dist} shows estimates of the distance to Tycho SNR as obtained
from the X-ray  flux normalization (column 2) and as derived from the brightness profiles (column 3). Also the value of 
distance to Tycho remnat from observed proper motion of filaments \citet{smith} presented.
The distance estimated from normalizations of the SNR spectrum is more than two times 
smaller than that derived from the brightness profiles. This may suggests that 
our model of Tycho SNR is underluminous.

\begin{table}
\caption{The estimates of the distance to Tycho SNR from W7 model, and estmimates of the distance from \citep{smith}}
\label{dist}
\begin{center}
\begin{tabular}{l|l|l}
\hline
\\
flux &  profiles & $H_\alpha$ filaments \\
\hline
\\
 $1.34 \pm 0.01$ kpc & $ 3.1 \pm 0.1$ kpc & $ 1.5 -3.1$ kpc\\
\hline
\end{tabular}
\end{center}
\end{table}

\section{Discussion}
\label{disc}
In this paper we have performed a hydrodynamical calculation of X-ray spectrum and 
radial brightness profiles in X-ray lines of SNR 1572 and compared the model 
X-ray emission  with the XMM-Newton observations.  We have used the classical deflagration 
model of thermonuclear supernova explosion W7 \citep{w7}. We employed the method presented in \elk\,
and complemented by the inner-shell collisional ionization processes in the calculations of X-ray emission 
from the remnant. This improvement allowed us to remove discrepancies 
between theoretical and observational data found in \elk.

We compared the model spectrum with the observable one by constructing the {\sc xspec} 
table model with 3 interpolation parameters: $q, C_\mathrm{kill}, \rho_\mathrm{CSM}$. This model was used to fit 
observational data and to find the best-fitting values of the interpolation parameters. 
The moderate value of the parameter  $C_\mathrm{kill} \simeq 0.01$ and moderate value of the parameter $q \simeq 0.9$
suggest that thermal conduction is not very important and points to an appreciable non-Coulomb energy exchange 
between electron and ion components. These values also suggest the presence
of magnetic field and plasma instabilities in the ejecta of the SNR.

The derived column density $N_H$ is slightly smaller than that estimated earlier by \citet{alb}, \citet{exosat}, \citet{itoh}, \citet{hwang} and
the value of $\rho_\mathrm{CSM}$ is higher than that derived by \citet{xmm}, \citet{hugh}, \citet{brink}.
Since it is not excluded that the W7 model is not the best one for the Tycho SNR, these
parameters should be considered as zero approximation. In forthcoming studies we plan to examine 
other models with different modes of burning in type Ia supernovae with the goal to select the best-fitting one.

Estimations of the distance to the remnant using the X-ray flux normalization and
remnant's angular size (Table~\ref{dist}) disagree likely due to neglecting
the nonthermal component from the forward shock  in X-ray emission modeling.
Furthermore, probably the explosion model of Tycho supernova was more energetic 
than W7 model.

\section{Conclusions}
\label{concl}
Our conclusions are as follows. The calculated X-ray spectrum and radial brightness profiles indicate that even the classical SN~Ia 
 deflagration model W7 is capable to reproduce the main observable features of a young SN Ia remnant, provided that all important physical 
processes are taken into account. In this case there is no need to introduce additional element mixing in 
the ejecta of this model \citep{itoh} 

\section*{acknowledgments}
The author thanks Blinnikov~S.I. and Postnov~K.A. for numerous helpful discussions 
and Barkov~M.V. and Immler~S. for remarks of material significance.
This work is supported by Russian Foundation for Basic Research (project nos. 
05-02-17480, 04-02-16720, 03-02-16110 and 03-02-16068).

\label{lastpage}
\end{document}